\newcommand{\subparagraph}{}

\documentclass[10pt, conference, compsocconf, letterpaper]{IEEEtran}
\usepackage{graphicx}
\usepackage{caption}
\usepackage{graphicx}
\usepackage{xcolor}
\usepackage{mathtools}
\usepackage{braket}
\usepackage{listings}
\usepackage{cite}
\usepackage{titlesec}
\usepackage{booktabs}
\usepackage{array}
\newcolumntype{L}[1]{>{\raggedright\let\newline\\\arraybackslash\hspace{0pt}}m{#1}}

\lstdefinelanguage{qasm}
{
  morekeywords={
    qubits,
    name,
    map,
    mov,
    cnot, h, x, y, x, not, cx, cz, cr, crk, prep_x, prep_z, prep_y, rx, ry, rz, s, t, sdag, tdag, measure, display, measure_x, measure_z, measure_parity,version
  },
  sensitive=false, 
  morecomment=[l]{\#}, 
  morecomment=[s]{/*}{*/}
}

\definecolor{eclipseBlue}{RGB}{42,0.0,255}
\definecolor{eclipseGreen}{RGB}{63,127,95}
\definecolor{eclipsePurple}{RGB}{127,0,85}

\def\note#1{}
\def\note#1{\textbf{\color{red}[#1]}}

\newcommand{\GATE}[1]{\texttt{\detokenize{#1}}} 
\newcommand\Mark[1]{\textsuperscript{#1}}

\begin{document}

\title{cQASM v1.0\\Towards a Common Quantum Assembly Language \\[1.25ex]
}




\author{\normalfont\large N. Khammassi\Mark{1}, G.G. Guerreschi\Mark{2}, I. Ashraf\Mark{1}, J. W. Hogaboam\Mark{2}, \\
    C. G. Almudever\Mark{1}, K. Bertels\Mark{1} \\
    \newline \\ \vspace{0.2cm}
    \Mark{1} QuTech, Delft University of Technology, The Netherlands \\ \vspace{0.2cm}
    \Mark{2} Intel Labs, USA \\ 
}





\maketitle

\begin{abstract}


The quantum assembly language (QASM) is a popular intermediate representation used in many quantum compilation and simulation tools to describe quantum circuits. Currently, multiple different dialects of QASM are used in different quantum computing tools. This makes the interaction between those tools tedious and time-consuming due to the need for translators between theses different syntaxes. Beside requiring a multitude of translators, the translation process exposes the constant risk of loosing information due to the potential incompatibilities between the different dialects. Moreover, several tools introduce details of specific target hardware or qubit technologies within the QASM syntax and prevent porting the code to other hardwares. In this paper, we propose a common QASM syntax definition, named cQASM, which aims to abstract away qubit technology details and guarantee the interoperability between all the quantum compilation and simulation tools supporting this standard. Our vision is to enable an extensive quantum computing toolbox shared by all the quantum computing community.\\

\end{abstract}



\section{Introduction}

Building quantum computers requires implementing multiple functional layers. At the most abstract level, algorithm designers formulate quantum algorithms
in a high-level language that requires one or more compilation steps to translate the algorithm description into a set of instructions that can be executed by quantum hardware. Compilers can internally use different intermediate representations to, for example, perform optimizations, instruction scheduling or qubits mapping, but it is desirable, for portability and flexibility, that the outcome of the compilation is a hardware-independent quantum assembly code (QASM). An additional translation process, possibly including further compilation and optimization steps, is then used to generate the hardware-specific micro-code. The micro-code can then be executed on the target micro-architecture which provides the hardware-based control logic needed to execute the instructions on the target quantum processor. \\

The quantum assembly language (QASM) is a hardware-independent instruction set which aims to provide a compact and expressive intermediate representation to describe quantum circuits. This representation is intended to be used not only by quantum compilers, but also quantum computer simulators, reversible circuit synthesis frameworks or microcode synthesis backends and other tools that needs to represent quantum circuits in a straightforward way. Algorithm designers can also benefit from a comprehensive QASM during the design and test of relatively small quantum circuits, algorithms and protocols. \\

QASM first appeared in 2005 in a set of software tools from MIT \cite{QASM-tools} and in \cite{qla}.  Since then, many different custom QASM dialects have been defined and used by various tools either as input or as an intermediate quantum circuit representation. This makes the interaction between the different tools difficult and time-consuming since it requires translations between different dialects with the constant risk of loosing information in the translation process when two dialects do not offer the same features or have different focus. With this paper, we propose the first version of a common QASM language (cQASM) to help bringing the quantum computing community together and trigger the effort of a standardized QASM language. Our goal is to provide a common ground for all the tools that have been or will be developed to compile, simulate and ultimately execute quantum circuits.\\

\begin{figure*}[h]
\includegraphics[scale=0.5]{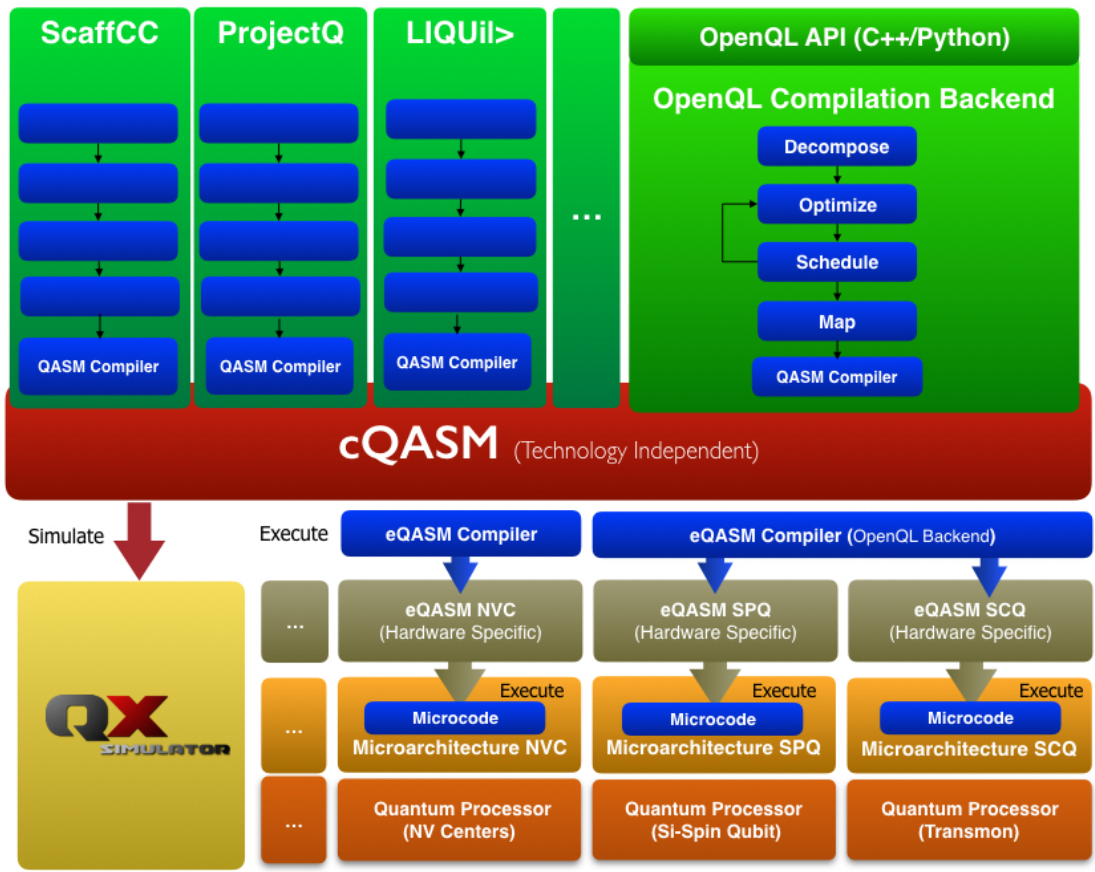}
\centering
\caption{Compilation of Quantum Algorithms : an algorithm written in high-level language can be optimized and compiled into a common technology-independent cQASM code then into hardware specific executable QASM (eQASM) code.}
\label{fig:stack}
\end{figure*}

In the definition of the cQASM syntax, we took particular care to include the user experiences and expectations from different parts of the quantum community. We discussed and included ways to describe the recent variational quantum algorithms and sampling tasks proposed for pre-error corrected devices, while we also defined compact commands to express active feedforward and error-correction related instructions. We expect that novel algorithmic solutions and the continuous progress in hardware architectures will continue to drive the development and adoption of new features. \\

Defining a common QASM guarantees the compatibility of the different tools and offers access to a large toolbox that includes tools for quantum compilation and simulation for the quantum computing community.\\

The work is organized as follows: in the next section we give a brief overview of some QASM variants discussed in the relevant literature of the past few years, while in section III we present the cQASM language. Its syntax definition is organized by topic according to:
\begin{itemize}
    \item Qubit Definition and Addressing
    \item Measurement Register
    \item Quantum Gates
    \item Quantum Circuit Definition
\end{itemize}

In section IV, we introduce special features that are optional extensions of the cQASM syntax addressing the requirements of specialized, but important, quantum computing tools or algorithms. Finally, Section V and IV give an overview of future extensions and draw our conclusions.

\section{Related Works}
Over the last few years, several QASM variants have been defined and used in various quantum computing tools.  The initial version of QASM was first introduced by Nielsen and Chuang to make the required illustrations for their book  \cite{NielsenChuangBook} and by Andrew Cross through a set of tools in \cite{QASM-tools}. QASM is now being used in different tools as the underlying quantum assembly language layer that translates the high level quantum algorithms in low level quantum operations which can then be executed either on quantum simulation platforms such as QX \cite{Khammassi2017} and QHipster \cite{Smelyanskiy2016} or on actual quantum processors such as demonstrated by QuTech and Intel on Superconducting and Si-Spin qubits \cite{Hipeac2017} or by IBM on the quantum experience platform \cite{IBMQE}. 
Different QASM dialects has been used by the different compilation and simulation tools such as the QASM defined by \cite{Khammassi2017} as input for QX or the OpenQASM defined by \cite{Cross2017a}.

As visually illustrated in Fig.~\ref{fig:stack}, we define cQASM as a hardware-independent language that is produced by quantum compilers as output of the translation process from a code written in high-level language into a quantum assembly code. Examples of compilation tools are ScaffCC \cite{ScaffCC15}, ProjectQ \cite{ProjectQ}, $LIQUi\ket{}$ \cite{liquid} and OpenQL \cite{openql}. These compilers can include a back-end pass to generate the technology-independent cQASM. \\

The cQASM instructions can then be used as input to a quantum computer simulator or as input to a lower-level compiler which generates hardware-specific instructions, or executable QASM (eQASM), suitable for execution by the target quantum processor. As a concrete example, the OpenQL compiler developed at QuTech supports multiple backends and generates different types of eQASM, such as the QuMis code introduced in \cite{MA17} and executable on a dedicated microarchitecture controlling superconducting qubits.



\section{QASM Definition}

\subsection{Case sensitivity and Comments}

The cQASM syntax is not case-sensitive, i.e. upper case letters are equivalent to lower case ones. To make the code more readable, the quantum programmer can add comments in his code. Comments start with ``\#" and can be added either in a separate line or at the end of a line containing code as shown in Code Example \ref{code:bellpair}.


\begin{lstlisting}[
caption=Creation of Bell state.,
label=code:bellpair,
float,floatplacement=H, 
belowskip=-1.0 \baselineskip
]
   version 1.0
   # define a quantum register of 2 qubits
   qubits 2
   
   # create a Bell pair via a Hadamard rotation
   h q[0]
   # followed by a CNOT gate
   # q[0]: control qubit, q[1]: target qubit
   cnot q[0],q[1]
   
   # measure both qubits to test correlations
   measure q[0]
   measure q[1]
\end{lstlisting}

\subsection{Qubit Definition}
\label{sec:qubit_definition}

\subsubsection{Qubit Register}
Qubits can be grouped into a simple quantum register and addressed by their index. A qubits register can be created by specifying the number of qubits as shown in Code Example \ref{code:bellpair} at Line 2.

\subsubsection{Qubit Addressing}


Once the number of qubits is defined, the qubits can be addressed individually through its default identifier ``q[i]'' where ``i'' is the identifier of the target qubit: if the quantum register contains N qubits, then $i\in\{0,\dots,N-1\}$ and the qubit identifiers are q[0], q[1], \dots, q[$N-1$]. As an example, observe line 5 of Code Example \ref{code:bellpair} where the Hadamard gate is applied to qubit 0.


\subsubsection{Measurement Register}

By default, a measurement register (binary register) is associated to the quantum register. It is mainly used to store the outcome of measurements: after measuring a qubit q[0], the result of the measurement is automatically stored into bit b[0]. The outcome b[0] can be the final result of some quantum computation or can be used to apply binary-controlled gates to other qubits.

\subsubsection{Naming Qubits and Measurement Outcomes}

In order to give a meaningful name to each qubit and make the quantum program more readable, it is possible to name qubits using the keyword ``map''. The instruction `map'' renames a single qubit according to its two arguments: the first is the current qubit identifier, the second is the additional name. The syntax is presented at line 5 of Code Example~\ref{code:renamequbits} in which qubit q[0] is named ``data'' according to the role played in the algorithm. Once the qubit has been renamed, ``data'' is equivalent to ``q[0]'' and can be used interchangeably.

\begin{lstlisting}[
caption=Renaming qubits and measurement outcomes and using binary controlled operations.,
label=code:renamequbits,
float,floatplacement=H, 
belowskip=-1.0 \baselineskip
]
   version 1.0
   # define a quantum register of 3 qubits
   qubits 3

   # rename qubits
   map q[0],data
   map q[1],ancilla
   map q[2],output

   # address qubits via their names
   prep_z data
   prep_z ancilla
   prep_z output
   cnot  data,ancilla
   cnot  data,extra
   
   # rename classical bit
   map b[1],error_syndrome
   measure ancilla
   
   #apply binary controlled Pauli-X gate
   c-x error_syndrome,q[2]
\end{lstlisting}

Similarly to the qubits, the measurement bits can be renamed using the ``map'' instruction to improve code readability.
In Code Example \ref{code:renamequbits}, qubit q[1] is renamed ``ancilla''. When q[1] is measured, the outcome is stored by default in b[1]. Such bit was previously renamed as ``error\_syndrome'' (at line 6). Its value can be used to (classically) control a Pauli-X gate which we apply to q[2] at the end of the circuit.

\subsection{Quantum Gates}

The QASM syntax supports an overcomplete, universal set of quantum gates which includes single-, two- and three-qubits gates, as listed in TABLE~\ref{gates}. It provides support to commonly adopted controlled gates such as CNOT and Toffoli gates. The syntax is common to all gates: the QASM instruction is followed by a space and a list of arguments separated by commas. The first arguments are the identifiers of the qubits involved in the gate (one identifier for single-qubit gates, two for two-qubit gates, \dots).
Certain gates require additional parameters: for example, single-qubit arbitrary rotations need to specify the value in radiant of the rotation angle. 

The set of gates is extended by allowing each gate to be binary-controlled. By prepending ``c-'' to the gate name, the quantum operation is executed if and only if a specific measurement outcome is equal to 1. In this case, the first parameter after the instruction name will not be the qubit identifier, but the identifier of the classical bit. Any gate defined in TABLE~\ref{gates} can be extended in the way just described.


\begin{table}[h!]
\centering
\caption{Supported Quantum Gates} 
\label{gates}
\begin{tabular}{|L{1.6cm} | L{3.4cm} | L{2.1cm} |}
\hline
\textbf{Quantum Gate}  & \textbf{Description} & \textbf{Example}  \\ \hline
I & Identity  &  i q[2]             \\ \hline
H & Hadamard & h q[0]             \\ \hline
X & Pauli-X & x q[1]             \\ \hline
Y & Pauli-Y & y q[3]             \\ \hline
Z & Pauli-Z & z q[7]             \\ \hline
Rx & Arbitrary x-rotation & rx q[0], 3.14      \\ \hline
Ry & Arbitrary y-rotation & ry q[3], 3.14      \\ \hline
Rz & Arbitrary z-rotation & rz q[2], 0.71      \\ \hline
X90 & X90 & x90 q[1]           \\ \hline
Y90 & Y90 & y90 q[0]           \\ \hline
mX90 & mX90 & mx90 q[1]          \\ \hline
mY90 & mY90 & my90 q[0]          \\ \hline
S & Phase & s q[6]             \\ \hline
Sdag    &Phase dagger &  sdag q[6]          \\ \hline
T & T & t q[1]             \\ \hline
Tdag & T dagger &  tdag q[3]          \\ \hline
CNOT & CNOT & cnot q[0],q[1]       \\ \hline
Toffoli & Toffoli & toffoli q[3],q[5],q[7] \\ \hline
CZ & CPHASE & cz q[1],q[2]         \\ \hline
SWAP & Swap & swap q[0],q[3]       \\ \hline
CRK &  Controlled Phase Shift ($\pi/2^k$ ) & crk q[0],q[1],k        \\ \hline
CR &  Controlled Phase Shift (arbitrary angle) & cr q[0],q[1],angle        \\ \hline
c-X & Binary-Controlled X & c-x b[0],q[2]        \\ \hline
c-Z & Binary-Controlled Z & c-z b[1],q[2]        \\ \hline
\end{tabular}
\end{table}

\begin{table}[h!]
\centering
\caption{Supported State Preparation and Measurement}
\label{prep-meas}
\begin{tabular}{|L{1.8cm} | L{3.4cm} | L{1.8cm} |}
\hline
\textbf{Instruction}  & \textbf{Description} & \textbf{Example}  \\ \hline
prep\_x  & State preparation in x basis   & prep\_x q[0]         \\ \hline
prep\_y  & State preparation in y basis   & prep\_y q[1]         \\ \hline
prep\_z  & State preparation in z basis   & prep\_z q[1]         \\ \hline
measure\_x  & Measurement in x basis   & measure\_x q[0]         \\ \hline
measure\_y  & Measurement in y basis   & measure\_y q[1]         \\ \hline
measure\_z , measure  & Measurement in z basis   & measure q[2]  \\ \hline
measure\_all  & Measurement of all qubits in z basis   & measure\_all \\ \hline
measure\_parity  & Measurement of the product of Pauli matrices & measure\_parity q[0],x,q[2],z  \\ \hline
\end{tabular}
\end{table}

\subsection{Measurements}
\label{sec:measurements}

\subsubsection{Partial Measurement (Single Qubit)}
As shown in the previous examples, each qubit can be measured in the z-basis individually using the keyword ``measure'' followed by the target qubit, for instance ``measure q[0]''. By default, the qubit is measured in the z-basis, the outcome is binary (either +1 or -1) and the probability of each of the two possibility is related to the probability of qubit q[0] being in state $\ket{0}$ or $\ket{1}$ respectively. At the end of the measurement, the state of qubit q[0] collapses into the corresponding z-basis state. Single qubit measurement in all three basis (x-, y-, z-basis) are allowed by using the dedicated instruction \GATE{measure_x}, \GATE{measure_y} or \GATE{measure_z}.


\subsubsection{Register Measurement (All Qubits)}
Alternatively, one can measure the entire quantum register at once using same keyword \GATE{measure_all} without specifying any target qubit.


\subsection{Feedback Support : Binary-Controlled Quantum Gates}
Binary-controlled gates are quantum gates which are controlled by measurement outcomes. The programmer can use a binary measurement outcome to control a quantum operation. The latter will be executed only if that binary value is 1. In the following example we put the state of the first qubit q[0] into superposition, then we measure it and use the measurement outcome b[0] to conditionally apply the Pauli-X gate on qubit q[1].


Multiple measurement outcomes can be used to control a quantum operation, in this case all the control bits are placed before the qubits used in that quantum operation.

\begin{lstlisting}[
caption=Multi-binary-controlled quantum gate.,
label=code:bincontrolled,
float,floatplacement=H, 
belowskip=-1.0 \baselineskip
]
   h q[0]
   measure_z q[0]  # measurement outcome in b0 
   # simple binary-contolled gate
   c-x b[0],q[1]   # apply Pauli-X to q[1] if b[0]=1
   measure_z q[2]
   measure_z q[3]
   measure_z q[4]
   # multi-binary controlled gate
   c-x b[2],b[3],b[4],q[5] # apply pauli-x to q4 if b2=1 and b3=1 and b4=1
   # binary controlled gate using an arbitrary mask :
   # we want to apply a Pauli-X to q[4] if b[0]=0 and b[1]=1  
   not b[0]             # negate b0
   c-x b[0],b[1],q[4]   # multi-bits controlled X gate
   not b[0]             # restore the measurement register 
\end{lstlisting}

Sometimes, the programmer might need to use an arbitrary binary mask where some measurement outcomes are ones and others are zeros. In this case the programmer can use the \GATE{not}€ classical operation to invert a bit before using it to control an operation.

\subsection{Display Results}
\label{sec:display}

In section~\ref{sec:measurements} we introduced the description of the measurement operations. After a measurement is performed, the outcome must be accessible to the user. This feature is implemented through the command ``display b[i]'' that returns the outcome value of the latest measurement involving the i-th qubit.

In addition, when the QASM code describes a simulation and not an actual experiment, more information about the quantum state may be readily accessible depending on the simulation tools. For example, simulators that represent quantum states as dense vectors store all the quantum amplitudes at each step of the algorithm. In this situation, the command \GATE{display} can be used to inspect the quantum state by printing all computational-basis amplitudes to the screen. \\


Later, Section \ref{sec:special-features} will introduce special features like the ``measurement averaging''. The command ``display'' is compatible with these extensions and should print the average measurement outcome of a qubit (a double precision float in [0..1]) along with the overall number of measurements (integer), the number of +1 and -1 measurements. The later information allows for more flexibility when post-processing the results.

\section{Special Features}
\label{sec:special-features}






\subsection{Parity Measurements (Pauli String Observables)}

Measurements do not always involve a single qubit at a time, in fact several important algorithms require multi-qubit measurements. This situation is fundamentally different from cases in which each qubit is measured independently. Here, we consider observables that are constituted by products of single-qubit Pauli matrices on a subset of qubits. In general, any observable can be described as a Pauli string after appropriate extra gates are applied.

When the measurement takes place at the end of the algorithm, as is the case for the important class of variational algorithms (see Code Example~\ref{code:variational}), two options are available.  The simplest strategy is to separately measure all qubits involved in the Pauli string and then multiply the measurement outcomes.

However, this strategy is not viable when the measurement must not extract more information then the single outcome bit. This is the case when parity measurements are performed during error correction codes or to extract information about stabilizer states.
The keyword ``measure\_parity'' is introduced to describe the latter scenario. It is followed by an even number of arguments: each pair of arguments are constituted by the qubit identifier and the specification of the corresponding measurement axis (chosen among X, Y, and Z axis). The number of qubits involved is automatically defined by (half) the number of arguments following the instruction. The outcome is a single bit that will be copied in all bit registers associated with the qubits involved in the measurement. See Code Example~\ref{code:parity} for the explicit use.

\begin{lstlisting}[
escapeinside={(*}{*)},
caption=Parity-like multi-qubit measurements.,
label=code:parity,
float,floatplacement=H, 
belowskip=-1.0 \baselineskip
]
   version 1.0
   # define a quantum register of 3 qubits
   qubits 4

   # apply a short sequence of gates
   h q[0]
   h q[1]
   h q[2]
   cnot q[2],q[3]
   
   # measure the Pauli string (*Z$_0$Z$_2$*)
   measure_parity q[0],z,q[2],z
   # the outcome is stored in both b[0] and b[2]
   
   # measure the Pauli string (*X$_1$Y$_3$*)
   measure_parity q[1],x,q[3],y
   # the outcome is stored in both b[1] and b[3]
\end{lstlisting}

\subsection{Demolition Measurement}

In certain hardware architectures, measuring one or more qubits not only collapses their state, but also removes the qubits from the register. Consider, for example, quantum linear optics setup with dual rail encoding: the qubit state is determined by a single photon being in one of two modes and the photon detection removes it from both modes leaving them empty. While it is important to properly describe this feature at the level of eQASM, for the scope of the current instruction set it is sufficient to assume that an extra qubit is added to substitute the demolished one. It is implicit in the instruction set that this extra qubit is renamed accordingly to the demolished one and that its state corresponds to the usual result of projective measurements.

\subsection{Parallel Quantum Gates}

A set of quantum gates can be scheduled to start in parallel using the syntax shown in Code Example \ref{code:parallelism}. Gates between brackets and  separated by ``$\mid$'' are scheduled to be executed in parallel. The brackets allow the expression of parallelism of a set of gates over multiple lines and avoid having verbose single lines. For instance, \GATE{prep_z} gates on q[0], q[1] and q[2] in Code Example \ref{code:parallelism} are parallel gates. Similarly, the measurements are scheduled to be executed simultaneously. 


\begin{lstlisting}[
caption=Parallelism Specification.,
label=code:parallelism,
float,floatplacement=H, 
belowskip=-1.0 \baselineskip
]
   { prep_z q[0] | prep_z q[1] | prep_z q[2] }
   h q[0, 1, 2]
   h q[0:2]
   cnot q[0], q[3]
   cnot q[1], q[3]
   { measure q[0] | measure q[1] | measure q[2] }
   c-x b[0:2],q[0]
\end{lstlisting}

\subsection{Single Gate Multiple-Qubits (SGMQ)}
In many cases, addressing many qubits at once can be very useful when applying a single quantum operation to a set of qubits similarly to SIMD (Single Instruction Multiple Data) fashion in classical computing. The following notations aim to simplify the addressing of a set of qubits:
\begin{itemize}
\item Contiguous range of qubits: a set of qubits within a contiguous range can be addressed as ``q[i:j]'', in this case the qubits \{i,i+1,$\dots$,j\} are included in the qubit set.
\item Arbitrary set of qubits: alternatively, an arbitrary set of qubits can be addressed as ``q[i,j,k,l]'' where i, j and k are arbitrary index within a valid range of qubits.
\item Arbitrary set of qubit ranges: finally the two previous addressing modes can be combined to match a set of contiguous ranges of qubits, for instance ``q[i:j,k:l,m:n] designates a set of 3 qubit ranges forming \{i,i+1,$\dots$,j\}$\cup$\{k,$\dots$,l\}$\cup$\{m,$\dots$,n\}.
\end{itemize}
In Code Example \ref{code:parallelism}, Line 2, a Hadamard gate is applied simultaneously to qubits 0, 1, 2. In case the indices are contiguous, the list can be abbreviated with ellipsis, e.g. h q[0:2] such as in Line 3 which is equivalent to line 2. \\

Similarly to the qubits, multiple measurement outcomes can be addressed simultaneously using the list notation the same way as the qubits, for instance, at the end of the previous, the measurement outcomes of qubits 0, 1 and 2 are used to control a \GATE{Pauli-X} quantum operation on a  qubit 0. This addressing mode reduce the verbosity of the code, improve its readability and offers a compact way to express parallelism.

\subsection{Sub-circuit Definition}

For better readability, a QASM circuit can be split into a set of functional sub-circuits. The list of QASM instructions forming a sub-circuit is provided, after the name of the sub-circuit itself, starting with a dot. Code Example~\ref{code:Grover} partition the circuit in 3 parts: the initialization starting at line 5, the single iteration of a Grover-step from line 17, and finally the measurements after line 37.

In addition, the definition of sub-circuits allow the specification of loops over each sub-circuit as detailed in the next paragraph and already included in example mentioned above.


\begin{lstlisting}[
caption=Grover Algorithm.,
label=code:Grover,
float,floatplacement=H, 
belowskip=-1.0 \baselineskip
]
   version 1.0
    
   # define a quantum register of 9 qubits
   qubits 9
   
   map q[4],oracle
 
    # sub-circuit for state initialization
    .init
        x oracle
        { h q[0] | h q[1] | h q[2] | h q[3] | h oracle } 
   
    # core step of Grover's algorithm
    # loop with 3 iterations
    .grover(3)

       # search for |x> = |0100>
       
       # oracle implementation
       x q[2] 
       toffoli q[0],q[1],q[5]
       toffoli q[1],q[5],q[6]
       toffoli q[2],q[6],q[7]
       toffoli q[3],q[7],q[8]
       cnot q[8],oracle
       toffoli q[3],q[7],q[8]
       toffoli q[2],q[6],q[7]
       toffoli q[1],q[5],q[6]
       toffoli q[0],q[1],q[5]
       x q[2]

       # Grover diffusion operator
       { h q[0] | h q[1] | h q[2] | h q[3] }
       { x q[0] | x q[1] | x q[2] | x q[3] }
       h q[3]
       toffoli q[0],q[1],q[5]
       toffoli q[1],q[5],q[6]
       toffoli q[2],q[6],q[7]
       cnot q[7],q[3]
       toffoli q[2],q[6],q[7]
       toffoli q[1],q[5],q[6]
       toffoli q[0],q[1],q[5]
       h q[3]
       { x q[0] | x q[1] | x q[2] | x q[3] }
       { h q[0] | h q[1] | h q[2] | h q[3] }
       display

   # final measurement
   .measure
       h oracle
       measure oracle
       display
\end{lstlisting}

\begin{table}[h]
\centering
\caption{Supported Code Constructs}
\label{const}
\begin{tabular}{|L{2.6cm} | L{2.65cm} | L{2.2cm} |}
\hline
\textbf{Statement}  & \textbf{Description} & \textbf{Example}  \\ \hline
\{gate args $\mid$ gate args\}  & Apply set of gates in parallel    & \{h q[0] $\mid$ h q[1] \}         \\ \hline
gate qubitlist  & Single gate multiple qubits operation    & h q[0,2,3]         \\ \hline
.function  & Functional sub-circuit   & (see Code Example \ref{code:Grover})        \\ \hline
.function(iterations)  & Static loops   &  (see Code Example \ref{code:Grover})  \\ \hline
\end{tabular}
\end{table}

\subsection{Static Loops}

The loop control structure is an essential component of many quantum algorithms such as the Grover's algorithm. For now, QASM provides support for the FOR-loop by simply adding the number of times a sub-circuit needs to be executed right after the name of the sub-circuit such as at line 10 of the Grover's algorithm example given in Figure 10.

\subsection{Quantum Gate Scheduling}

Instruction scheduling in the VLIW fashion requires the specification of ``bundles'' which can be expressed using a \GATE{wait} instruction (example: $wait\,5$ specifies that an execution unit should wait 5 time units before executing the next instruction). Together with the parallelism specification, the \GATE{wait} instruction provides accurate time specification without changing the structure or the syntax of the QASM language. \\

Beside setting the starting time of each instruction in a scheduled QASM code, the ``wait'' instruction effectively sets the duration of each instruction. This possibility allows the user to determine by hand the duration of each kind of gate and, in simulation mode, allows to tailor the simulation to different qubit technologies characterized by different gate durations. The gate duration is a critical information in many cases such as noise simulation where qubits suffer decoherence over time or quantum computer architecture simulation where the overall execution time of a quantum program needs to be calculated. \\

Should be noted that the simulators or tools which do not support the notion of the time information can ignore the ``wait'' instruction, the duration of each instruction is then a single cycle. The number of cycles is an integer equal to or greater than 1. 


\subsection{Measurement Averaging}
In many quantum algorithms, the measurement outcomes of the qubits are not directly used as raw binary values, but collected and post-processed. One common form of post processing consists in computing the average measurement outcome of a given qubit. For instance, a programmer might need to evaluate the robustness of an error correction code through characterizing the physical vs logical error rate, for that the quantum circuit is executed several times under quantum noise and the average measurements are collected and used to compute the actual error rate. Another example is the execution of experiments to observe Rabi oscillations.\\

The ubiquitous necessity of this feature makes it an attractive candidate for implementation in simulation tools and in actual hardware. In such case the measurement outcomes for each qubit are accumulated and the average ground state measurement is stored and returned when requested. For instance, a programmer can execute a loop with several thousands of iterations and the measurement are collected and made available at the end of this loop. The available information includes the number of +1 and -1 measurements for each qubit and by extension the average measurement in the ground or excited state. The \GATE{display} command described in section~\ref{sec:display} can display this information at any point of the quantum circuit.  \\

In order to accommodate such use cases, one syntax element allowing the resetting of the average measurement for each qubit when needed is required: a command named \GATE{reset_averaging} can be used to reset the averaging engine whenever needed. The syntax of the later command is as follows: \GATE{reset_averaging q[i]} where q[i] is one or more qubits identifier such as specified in the qubit addressing Section~\ref{sec:qubit_definition}.

\subsection{Expectation Value of Observables}

\begin{lstlisting}[
caption=Central routine of a typical variational quantum algorithm. Here we prepare a certain quantum state and then measure the expectation value of the observable ${\hat{A} = \alpha Z_1 + \beta X_0 X_2}$.
%\note{GG: the manipulation of the outcome statistics comes more natural in QASM 2.0 so I suggest to include this user-case in limitations and extensions. This example can be included in the QASM 2.0 manuscript.}
,
label=code:variational,
float,floatplacement=H, 
belowskip=-1.0 \baselineskip
]
    version 1.0
    # define a quantum register of 3 qubits
    qubits 3

    # reset the counters for the average procedure
    reset_averaging

    # prepare and measure the quantum state 1000 times
    # to accumulate a large outcome statistics
    .average(1000)
        # state preparation
        prep_z q[0:3]
        { rx q[0] 3.14 | ry q[1] 0.23 | h q[2] }
        rx q[2] 3.14
        cnot q[2],q[1]
        { z q[1] | rx q[2] 1.57 }

        # measure of $Z_1$
        measure_z q[1]
        # the corresponding average is automatically updated

        # measure of $X_0 X_2$
        measure_parity q[0],x,q[2],x
        # the corresponding average is automatically updated

    # estimate the observable A
    .result
        # show the average of $X_0 X_2$ together with its latest outcome
        display b[0]
        # show the average of $Z_1$ together with its latest outcome
        display b[1]
        
    # the expectation value of $\hat{A}$ follows
    # from a straightforward postprocess 
\end{lstlisting}

Recently, a novel class of quantum algorithms emerged that is particularly relevant to near term devices due to its intrinsic robustness to systematic noise. These algorithms are commonly called \emph{variational quantum algorithms} \cite{Farhi2014,Peruzzo2014,VQA} and alternate short quantum computations with classical post-process of the outcomes: the basic idea is that a short quantum circuit suffices to prepare an approximate solution of the problem at hand, but the specific form of the quantum gates is \emph{a priori} unclear and must be determined by try-and-fail approach guided by a classical optimizer. To implement this kind of hybrid algorithms, one needs to 1) compute expectation values of several, possibly non-commuting, observables, 2) compute the quality of the approximate solution, 3) provide this information to a classical optimizer that suggests how to modify the quantum circuit to improve the approximate solution, and 4) reprogram the quantum computer before start the next iteration from point 1). 

While the development of instructions for a natural execution of the quantum-to-classical-to-quantum iteration is outside of the scope of the current work, we believe that it is important to illustrate how one can perform the steps 1) and 2) above using the instructions already introduced. Code Example~\ref{code:variational} uses static loops to prepare and measure 1000 times the relevant terms of observable $\hat{A}$. The average is obtained through the dedicated command introduced in the previous subsection.


\section{Future Extensions}
The proposed QASM syntax specification allows the description of quantum circuits in a relatively compact way while remaining at the quantum gate level and abstracting away low-level hardware details. Despite being an assembly language, several features such as SGMQ instructions allows the expression of gate-level parallelism in a relatively compact way and allowing simulation tools and compiler to specify the execution timing and the gate scheduling in a straightforward way. \\

Despite the expressiveness that can be provided by the proposed cQASM syntax, several limitations worth to be mentioned and addressed in future versions: 
\begin{itemize}
    \item Control flow and conditional branching.
    \item Interleaving classical and quantum instructions: this requires classical registers and a classical instruction set.
    \item Sub-circuit reuse: currently, the sub-circuits cannot be recalled many times, the reuse of a sub-circuit as a recallable function can offer make algorithm specification even more compact, promote code reuse and save instruction space.  
    \item Along with classical instructions,  a classical memory can offer a storage space for intermediate classical computations which can be needed in some quantum algorithms.
\end{itemize}



\section{Conclusion}

In this paper we proposed the common Quantum Assembly language (cQASM) to pave the way toward a hardware-agnostic language that is shared by the quantum computing community and implemented by the many tools required to achieve large-scale quantum computation. We described the cQASM syntax and the semantic behind it and we distinguished two main parts of the syntax, one minimal syntax required to express basic quantum circuits, and a set of extensions to offer different features such as parallelism expression, loop constructions or special measurements.\\

The first version of cQASM allows the intuitive description of quantum circuits, but its current syntax does not address the interaction between quantum and classical computation in a completely satisfactory way. These important features, like branch support, conditional loops or dynamic rotation angle computation, will be introduced in future version where classical instructions can be interleaved with quantum ones and interact together. We hope for a large participation in the definition of the instruction set regulating the interface between quantum and classical aspects of quantum algorithms and their execution.\\

\section*{Acknowledgment}

The authors would like to thank Prof. F. Chong, Y. Ding and A. Holmes (University of Chicago), Dr. A. J. Abhari (IBM), Prof. M. R. Martonosi (Princeton University) for their contributions and valuable discussions in defining the syntax of the common quantum assembly language v1.0.\\
\bibliographystyle{IEEEtran}
\bibliography{references}

\end{document}